\documentclass{PoS}

\title{Are the modern computer simulations a substitute for physical models?  The SKA case.}

\ShortTitle{The Murchison Widefield Array}

\author{\speaker{S.J. Tingay}\\
        International Centre for Radio Astronomy Research - Curtin University, Perth, Australia\\
        E-mail: \email{S.Tingay@curtin.edu.au}}

\abstract{I consider the question posed to me by the scientific organisers of the conference, ``Are the modern computer simulations a substitute for physical models?  The SKA case.''  I briefly consider the current knowledge of computer simulations and of physical prototypes in the context of understanding interferometric radio telescopes.  My conclusion is that, ``no, computer simulations are not a substitute for physical models when it comes to understanding the SKA.....furthermore, physical models are not much help either.''  This conclusion is intentionally provocative, designed to promote some discussion at the conference, which it did.  However, the conclusion reflects my belief that we do not have a deep enough understanding, theoretical or practical, of how interferometry works, to determine if the SKA will meet the stated specifications or not.  I conclude that we need to adopt a qualitatively different approach to dealing with interferometric data.  I note that some good work is being done on this front, but it is likely a bigger effort is needed in the SKA era.  This is exactly the type of innovation that projects such as the SKA should encourage.}

\FullConference{From Antikythera to the Square Kilometre Array: Lessons from the Ancients\\June 12-15, 2012\\
		Kerastari, Greece}

\begin{document}

\section{Introduction}
This contribution to the proceedings summarises my oral presentation, of the same title, at the conference.  My presentation was one of a pair, with the other titled ``Are the modern computer simulations a substitute for physical models?  The Antikythera case'', by Manos Roumeliotis.  The intention of these connected presentations was to explore the ways that computer simulations of complex systems, for example the SKA and the Antikythera device, can be used to further our understanding of how these systems could be expected to work, and to ponder whether computer simulations are better than constructing physical (hardware) models or prototypes.

I suppose Ron Ekers, as Chair of the Scientific Organising Committee, asked me to give this presentation as I've historically had a foot in both hardware prototyping and computer simulation camps.  For a few years I was Chair of the SKA Configurations Simulations Task Force, charged with producing SKA computer simulations that ultimately helped choose two sites from the four candidates, in 2006.  Conversely, I'm Director of the Murchison Widefield Array (MWA\footnote{Updates and news items for the MWA project are regularly posted at:\\ http://www.facebook.com/Murchison.Widefield.Array}), which is a substantial technology and science precursor for the low frequency SKA \cite{ska,skascience} being built in Western Australia \cite{tin12, Lonsdale-etal.2009}.  Thus, I was suitably torn when I came to prepare for the presentation.  The question posed caused me to think about the pros and cons of computer simulations and hardware prototyping and, specifically, where the radio astronomy community sits in terms of understanding the nature of interferometric telescopes with two orders of magnitude more collecting area than those currently being used.

I do not claim that the following is a complete or comprehensive treatment of the issues.  In a couple of places I purposely do not make lists of failed endevours.  This article is more a statement of my broad opinion, intended to promote discussion, than an in-depth review of the field.

\section{Goals of modelling (computer and/or physical)}

The aims of any type of modelling for the SKA, be it via computer simulations or hardware prototypes, are usually to assess the expected performance of the system, relative to a set of science goals and technical requirements.  The standard (although not always followed) project-based approach is to define a set of science goals for a new instrument.  These could be based on, or motivated by, emerging theoretical or observational evidence that points toward fundamental questions in astrophysics that require answers.  The science goals should then be converted into technical requirements i.e. how would a system be implemented that could address the science goals?  Finally, when the technical requirements have been formulated, the project scope can be defined, costed and scheduled.  This process is not linear and should be considered iterative, as an initial set of technical requirements may eventually be judged as not feasible.  Or perhaps once costed and scheduled, a particular implementation may be judged to cost too much or take too long to build.  Iteration is required in order to loop back to science goals or technical requirements and make trade-offs that reduce cost, complexity, time, or some other parameter that makes the project feasible.  Conversely, in the rare (unknown?) event that the final plan costs less than expected, additional science goals or stricter technical requirements could be added to increase the scope of the instrument.

An illustrative example I used in my conference presentation was the metric of imaging dynamic range.  This metric describes a technical requirement of the SKA, that the dynamic range in the images of the sky be in excess of one million to one.  The dynamic range metric describes the ratio of the brightest astrophysical feature in the image to the most prominent artifact in the image.  This SKA technical requirement aims to push radio source counts $\sim$100 times deeper than possible with the best previous instrumentation, the NRAO Very Large Array (a science goal).  Why is such a large dynamic range required?  Because the SKA will contain so much collecting area, it will be very sensitive and it should be capable of detecting very faint objects.  However, it is typically the case that the presence of very bright objects in the image produces artifacts distributed around the image.  If the artifacts are brighter than the weakest sources detectable, then differentiating between real objects and artifacts can be impossible.  The million to one dynamic range for the SKA is therefore a function of the raw sensitivity of the instrument, the brightest object expected in each field, and the calibratibility of the instrument.  The latter depends on a wide range of instrument parameters, hardware choices and calibration algorithms.  Thus, from a back of the envelope calculation based on relatively simple considerations of radio source populations, a relatively simple technical requirement can be stated.  However, determining if this technical requirement is feasible or not is a highly complex undertaking.  Models, analytic, numerical or physical need to be developed to explore trade-offs that affect the metric.

The framework in which trade-offs are considered is therefore a fundamental and critical aspect of managing large and complex projects.  Prototyping or computer simulations are two methods for exploring trade-off space and considering questions of performance and cost in the context of the science goals and/or technical requirements, in a controlled fashion.  Both methods have their pros and cons.

\subsection{Computer simulations: pros and cons}

On the plus side of the ledger, computer simulations of systems such as the SKA can:

\begin{itemize}

\item isolate particular parts of the problem - focusing on identifying possible show-stopping effects;

\item once set up, allow many different simulations quickly.  This allows a systematic exploration of parameter space to reveal dependencies in the system;

\end{itemize}

On the other hand:

\begin{itemize}

\item the simulation output is only as good as the simulation input (as my year 11 computer science teacher said, ``rubbish in, rubbish out'');

\item the simulation of large and complex systems such as the SKA requires as much compute (or more) as the operational instrument itself.  For the SKA, this is well into the peta-scale or exa-scale.

\end{itemize}

\subsection{Physical models: pros and cons}

The benefits and advantages of physical hardware prototypes are:

\begin{itemize}

\item similar to computer simulations, one can isolate particular parts of the problem - focussing on show-stopping effects;

\item Whereas computer simulations can only deal with idealised systems, physical hardware prototypes incorporate subtleties and effects that are hard to simulate.  For example, digital hardware systems so critical to modern radio telescopes are supposed to operate on binary logic, however the design and programing of these systems is never so straightforward.  Physical prototypes can therefore uncover real-world issues that computer simulations cannot.

\end{itemize}

Conversely:

\begin{itemize}

\item physical prototyping can take a long time and be very expensive;

\item similar to computer simulations the results are only as good as the design inputs and the implementation.

\end{itemize}

\section{Conclusion \#1}

From the very brief list of pros and cons above, it should be clear that computer simulations and hardware prototyping can bring different advantages and disadvantages and should be applied in different situations.  In looking around at the suite of successful computer simulations of large-scale radio telescopes, it is clear to me that these efforts do not emerge from a vacuum, they come from groups and individuals with deep and practical experience of building and operating telescopes over a long period of time.  Examples of these successes are the ASKAP \cite{jon08,jon07} simulations of (primarily) Matthew Whiting and Tim Cornwell at CSIRO\footnote{http://www.atnf.csiro.au/people/Matthew.Whiting/ASKAPsimulations.php} or the Meqtrees\footnote{http://www.astron.nl/meqwiki} development of (primarily) Oleg Smirnov and Jan Noordam connected to LOFAR \cite{lofar} at ASTRON.  

Thus, effective computer simulations and hardware prototyping go hand-in-hand for best effect, and are generated by groups with the wisdom and the knowledge regarding where to deploy each effort for maximum gain.

The converse is that the worst computer simulation efforts do emerge from a vacuum, by turning the handle on endless simulations without the wisdom and knowledge to know what the simulations are revealing or what the appropriate questions are.  I'll choose to not list these efforts here.

\section{Conclusion \#2}

And so, to the main question: are the modern computer simulations a substitute for physical models when it comes to the case of the SKA?

My initial tendency when thinking about the answer to this question, given my background and experience with both physical and computer models, was to be diplomatic and eclectic.  I thought, was it possible that the combination of the two approaches offered a complete and complementary set of techniques that allowed a full solution to the challenges of understanding the SKA?  I found that I could not really back this statement up in a rigourous manner and that it was a fairly unsatisfactory answer in terms of the aims of the conference, which were to promote discussion and hopefully stimulate some new ideas.  So, I think a more provocative and ambitious answer is required.

This led me to consider an answer in the definite affirmative.  Yes, computer simulations are enough to understand the SKA.  After all, the physics of electromagnetic waves and their interactions with matter are well understood.  The equations of interferometry are relatively simple and also well understood.  Finally, massive amounts of computing time are available to tackle extremely complex simulations, such as in climate modelling.  So, essentially all the elements are in place, from the understanding of physics, algorithms and high performance computing.  So, why not simply yes?

Well, reality gets in the way.  Do we really understand these aspects of the problem as well as we think or as well as required to understand the SKA?  And is the relatively small radio astronomy community really positioned to tackle computer simulations on the scale that climate models do?  So, I also had to consider an answer of definite no, computer simulations cannot replace physical models, in practice.  

At this point I realised the problem with the question posed to me for the presentation - ``Are the modern computer simulations a substitute for physical models?  The Antikythera case''.  Implicit in the question is the assumption that we can actually understand the SKA through the construction of physical models, that physical models somehow provide a gold standard that simulations can only hope to challenge.  The burden placed upon simulations is to perform as well as physical models, with no expectation of scrutiny of physical models.  The problem I see is that, because of our collective experience with previous instruments, we tend to have an in-built bias in the way we consider prototyping, a bias toward physical models.  This subtle bias is reflected in the question.  Once I recognised the bias and the problem with the question posed to me, I was led me to my final conclusion.  I realised that a seriously critical look at both physical prototyping and simulations is required, that the burden of proof is not solely on simulations, as the question implies.  The answer to the question posed to me is: no, computer simulations are not a substitute for physical models when it comes to understanding the SKA.....furthermore, physical models are not much help either.

Right, so I'm going beyond my brief and I'm saying that we are not really in a position to understand how the SKA will work, at the stringent levels required to meet the SKA technical specifications as currently stated, via any route.  Ok, that's a fairly provocative statement (which did promote some discussion at the conference!).  What is the reasoning here?

An understanding of how interferometric radio telescopes work can be expressed in different ways, depending on how you approach the task.  If you are using a computer model, that understanding needs to be expressed in how you construct your simulation, encoding all the processes that map the inputs into the system onto a realistic set of outputs, to build a virtual SKA in software.

If you are building a physical model, you have a set of hardware that maps the inputs onto a set of outputs (measurements).  In order to interpret the outputs and understand how the hardware is doing the mapping, you need to apply your understanding of the instrument by calibrating the outputs.

Thus, {\it simulation} and {\it calibration} encode exactly the same understanding of interferometry in different ways.  If you have enough knowledge to calibrate the SKA, you have enough knowledge to simulate the SKA.  So, the key thing for me to consider is, how well do we understand the calibration of the largest and most complicated radio interferometers that currently exist?  These existing instruments, be they the JVLA, LOFAR, the ATCA, GMRT or WSRT could be considered the most advanced physical models we have for the SKA.  Are the data from these instruments routinely calibrated to levels that correspond to their theoretical peak performance?  Therefore, is our knowledge of interferometry good enough to understand these instruments as physical models for the SKA?  I think the answer would have to be no.  

I will not single out a list of examples of these failures (even though I did in my presentation), because they represent extraordinary efforts by extremely clever people to get the best out of our instruments, and the results are still easily good enough to undertake incredible astrophysics.  However, we cannot escape the fact that, in an absolute sense, we do not get the best out of instruments that have been around for a long time and have been exercised extensively by the best in the business; they still fall short of what they should be able to achieve.

It is true that for some instruments, in very special and optimised situations, with a lot of hands-on work by gurus, some results have come close to achieving what should be possible in terms of imaging and calibration.  However, this is quite a long way from having an instrument routinely producing these results in general circumstances, without a lot of hand-holding and manual intervention.  What magnifies the challenge for the SKA are the massive data rates and volumes involved, which are many orders of magnitude beyond what a human or team of humans, no matter how expert, can deal with.  The SKA will require the automatic application of highly accurate calibration algorithms to vast volumes of data in real-time.  The calibration algorithms have to therefore be generally applicable and highly robust, as well as accurate.

It feels to me that we need to find qualitatively new ways to deal with data from interferometric radio telescopes, that put the analysis and calibration closer to the domain in which the data are measured, and that this may lead to the understanding required for instruments as complex as the SKA.  Rather than just being provocative on this front, I'm putting my money where my mouth is.  The Murchison Widefield Array (MWA) is a low radio frequency interferometer using aperture arrays, with a very wide field of view \cite{tin12}, exposing a very wide range of issues in interferometric calibration and imaging.  The MWA is adopting a real-time imaging and calibration pipeline that attempts to do as much calibration and analysis in the measurement plane as possible, before going to the image plane \cite{mitchell}.  Also, one of my very clever postdocs is making good progress constructing a new framework for understanding interferometric data and calibration in the measurement plane, by merging the equations of interferometry with the fundamental underpinnings of Information Theory, with application to a range of astrophysical problems such as transient detection and detection of the Epoch of Reionisation \cite{trott1,trott2}.

So, I think that a range of solutions need to be explored and we should not feel too bound by the traditional approaches used for current instruments that are very small and limited compared to the SKA.  I also feel certain that radio astronomers would do very well to look at other disciplines to see what we can learn.  The conference on Antikythera and SKA was a great opportunity to start some very interesting cross-disciplinary discussions on these topics and I hope they continue in the future.

\section*{Acknowledgments}

I would like to thank Dr Tasso Tzioumis, for organising another excellent conference in his beautiful home village and Prof. Ron Ekers for coming up with the concept for the conference. My attendance was supported by the International Centre for Radio Astronomy research (ICRAR) and I acknowledge support from the Western Australian Government, through their Western Australian Premier's Fellowship program.

\end{document}